 \documentclass{camera}

\usepackage{graphicx} 

\def\fref#1{Fig.\,\ref{#1}}
\def\etal{et al.}
\def\lnA{\langle\ln A\rangle}
\def\Xmax{X_{\mbox{\scriptsize max}}}
\def\line{---}
\def\dashed{-\,-\,-}
\def\dotted{$\cdot\cdot\cdot$}
\def\hh{\vspace*{-4mm}}

\newcount\flag
\input epsf

\begin{document}

%

\title{THE COMPOSITION OF COSMIC RAYS AT THE KNEE
       \footnote{Invited talk,
       presented at the Workshop on Frontier Objects in Astrophysics and
       Particle Physics, Vulcano, May 24$^{th}$ - $29^{th}$, 2004.}
      }

%
\author{J\"ORG R. H\"ORANDEL}

%

\organization{University of Karlsruhe, Institut f\"ur Experimentelle Kernphysik,
PO~Box~3640, 76021 Karlsruhe, Germany --- www-ik.fzk.de/$\sim$joerg}

\maketitle

\begin{abstract}
The present experimental status concerning the composition of cosmic rays in
the PeV region is reviewed. The results are compared to predictions of
contemporary models for the acceleration and propagation of galactic cosmic
rays.
\end{abstract}

%
\section{Introduction}
\begin{figure}
 \includegraphics[width=0.59\columnwidth]{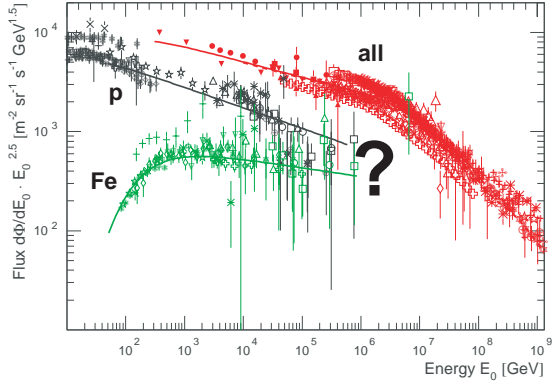}
 \begin{minipage}[b]{0.4\columnwidth}
   \caption{Energy spectrum of cosmic rays for all particles (dark grey),
   protons (black), and iron nuclei (light grey), for references see
   \cite{knie}.}
   \label{espek} 
 \end{minipage} \hh
\end{figure}

The solar system is permanently exposed to a vast flux of highly energetic and
fully ionized atomic nuclei, the cosmic rays. Their energies extend from the
GeV range to at least $10^{20}$~eV.  Over a wide range the energy spectrum
follows a power law $dN/dE\propto E^\gamma$. The spectral index changes around
4~PeV from $\gamma\approx-2.7$ to $\gamma\approx-3.1$. This transition
phenomenon has been reported for the first time in 1958 \cite{kulikov} and is
commonly referred to as {\sl the knee}.  A cutout of the energy spectrum in the 
knee region is depicted in \fref{espek}. Shown is the all-particle energy 
spectrum, which exhibits the knee, as well as results from direct measurements 
above the atmosphere for primary protons and iron nuclei. 

The origin of the knee is one of the central questions of high-energy
astroparticle physics, closely related to the mechanisms of acceleration and
propagation of high-energy cosmic rays. Answers are expected from the
measurements of the energy spectra for individual elements (or at least
elemental groups) above 1~PeV. Since this is an extreme experimental challenge,
one often measures the average atomic mass instead.

In this article, the present experimental status is reviewed and the results
are set in context to predictions of contemporary models for the acceleration
and propagation of cosmic rays.

Ideally, one would like to continue the energy spectra for individual elements
above several $10^{14}$~eV with direct measurements above the atmosphere. At
present, several groups measure in this region, applying different experimental
techniques, e.g. a calorimeter (ATIC \cite{atic}) or a transition radiation
detector (TRACER \cite{tracer}).  The TRACER experiment, investigating nuclei
from oxygen to iron, had a 14 day flight from Mc Murdo, Antarctica in December
2003.  Recent accelerator tests have shown that transition radiation detectors
can be utilized to measure energy spectra of cosmic rays in a space borne
experiment up to energies of about 1~PeV/n \cite{cerntrd}.

Presently, at energies above 1~PeV one relies on indirect measurements.  In
these experiments the secondary products, generated by interactions of
cosmic-ray particles with nuclei in the atmosphere are investigated.  Two basic
approaches can be distinguished: Measuring the debris of the particle cascade at
ground level by registering electrons, muons, or hadrons. Or measuring the
longitudinal shower development in the atmosphere by exploration of the
\v{C}erenkov or Fluorescence light generated predominantly by the shower
electrons.  An astrophysical interpretation of air shower data requires
detailed knowledge of the interaction processes in the atmosphere. One
of the tasks of air shower experiments is to improve the understanding of
high-energy interactions above the energies covered by todays accelerator
experiments and beyond their kinematical bound, see e.g. \cite{hadron}.

\section{Origin of the knee}
The bulk of cosmic rays is assumed to be accelerated in shock fronts of
supernova remnants (SNRs). This goes back to an idea of Baade and Zwicky, who
have estimated the total power required to generate the observed flux of cosmic
rays \cite{baade}. It can be shown, that about 3 supernovae per galaxy and
century are sufficient to release enough kinetic energy in order to deliver the
required power. A mechanism to accelerate particles by moving magnetic clouds
has been introduced by Fermi \cite{fermi}. The present understanding of
acceleration in strong shock fronts has been initiated by Blanford and Ostriker
\cite{blanford} which could demonstrate that at strong shocks particles are
accelerated efficiently.  The finite lifetime of a shock front ($\sim
10^5$~a) limits the maximum energy attainable to $E_{max}\sim Z\cdot (0.1 -
5)$~PeV for particles with charge $Z$.  Various versions of this scenario have
been discussed, see e.g. 
\cite{berezhko,stanev,kobayakawa,sveshnikova,wolfendale}.
In the literature also other possibilities, like the acceleration of particles
in $\gamma$-ray bursts are discussed \cite{plaga,wick}.

After acceleration, the particles propagate in a diffusive process for about
$20\cdot10^6$~a through the Galaxy, being deflected many times by the randomly
oriented galactic magnetic fields ($B\sim3$~$\mu$G).  The propagation is
accompanied by leakage of particles out of the Galaxy. With increasing energy it
is more and more difficult to magnetically bind the nuclei to the Galaxy. The
pathlength of traversed material decreases as $\Lambda\propto E^\delta$, with
$\delta\approx-0.6$.
Many approaches have been undertaken to describe the propagation process,
see \cite{ptuskin,ogio,roulet,swordy,lagutin}.
During the propagation phase, reacceleration of particles has been suggested at
shock fronts originating from the galactic wind \cite{voelk}.

Both, the acceleration and the propagation/leakage processes are expected to
yield cut-offs for the fluxes of nuclei at energies proportional to their
nuclear charge $E_k\propto Z$.

As further possible causes for the knee, interactions of cosmic-ray nuclei with
background photons or neutrinos in the Galaxy \cite{tkaczyk,dova,candia} or new
types of interactions in the atmosphere \cite{kazanas} have been discussed.
For a more comprehensive review of models the reader is referred to
\cite{origin}.

\begin{figure}
 \includegraphics[width=0.4\columnwidth]{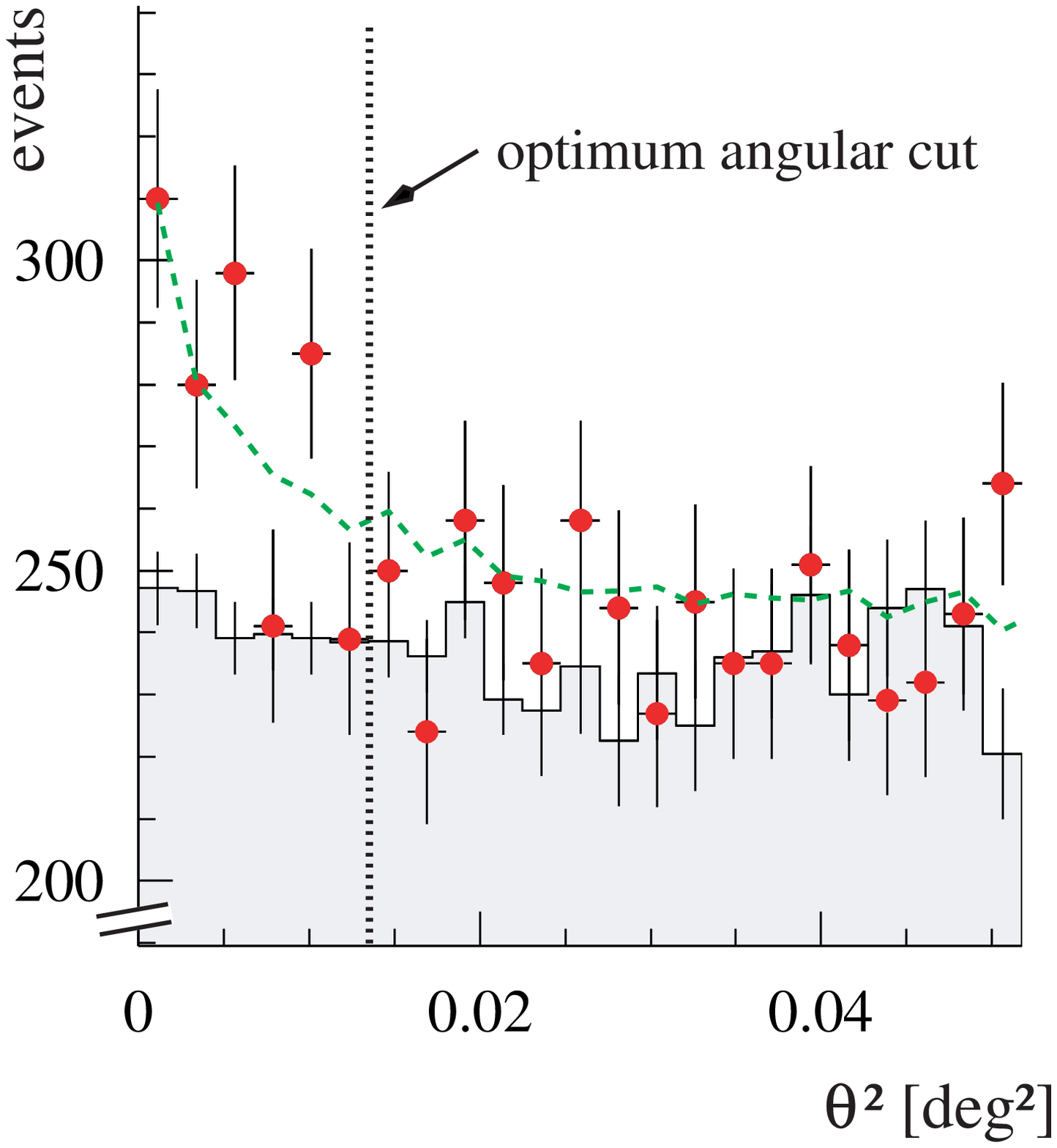}
 \hspace*{\fill}
 \includegraphics[width=0.5\columnwidth]{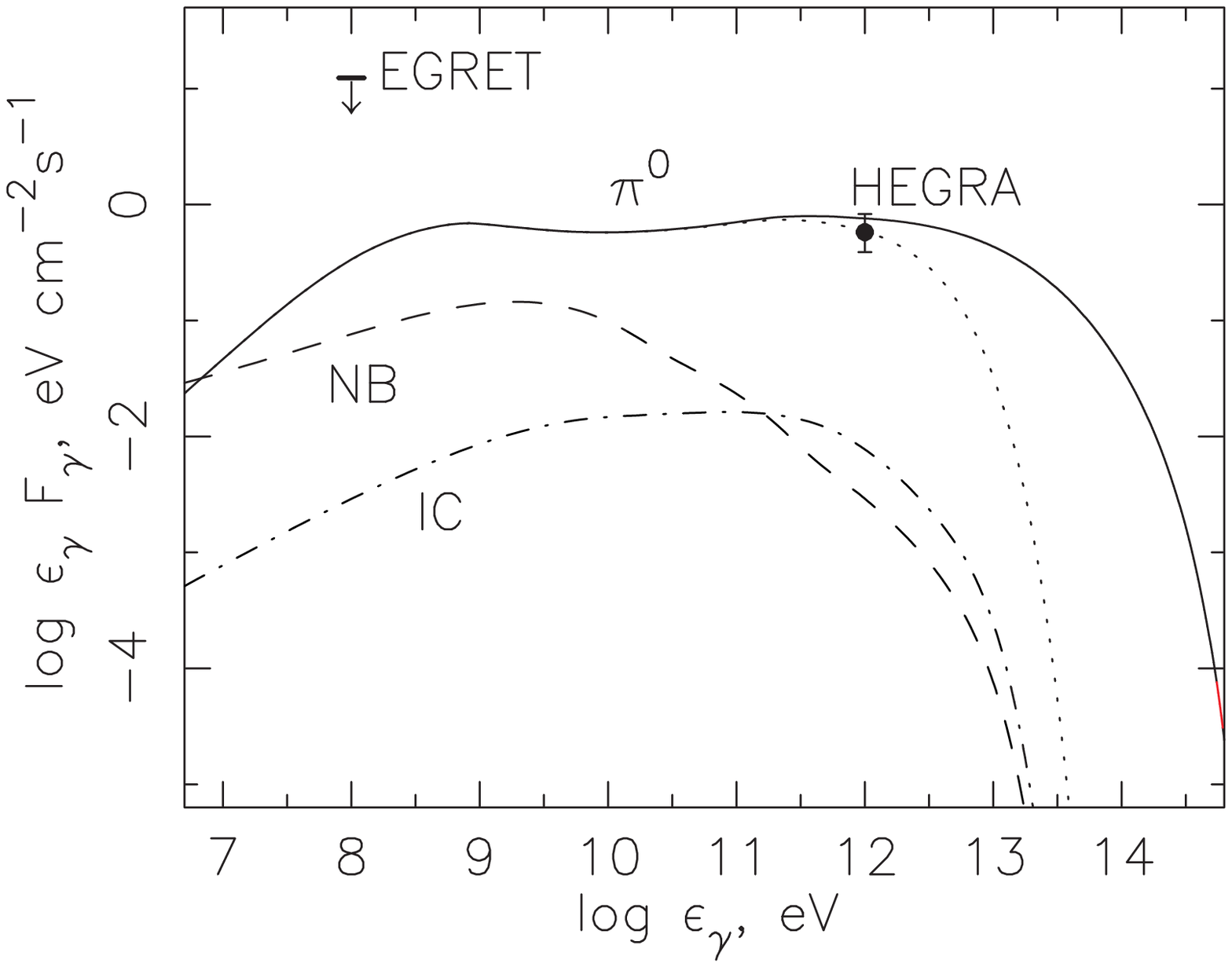}
 \caption{Left: The dots indicate the measured $\gamma$-ray signal as function 
  of angular distance to Cassiopeia A as measured by the HEGRA experiment. The
  shaded area is a background estimate \cite{hegra}.  Right: Integral
  $\gamma$-ray fluxes due to inverse Compton scattering, non-thermal
  bremsstrahlung and $\pi^0$-decay according to the model by Berezhko \etal\
  compared with measurements of the HEGRA and EGRET experiments
  \cite{berezhkocasa}. The dotted line is obtained for an assumed cut-off at
  4~TeV.}
 \label{casa}  \hh
\end{figure}

Electromagnetic emission of SNR has been detected in a wide energy range from
radio wave lengths to the x-ray regime. The observations can be interpreted as
synchrotron emission from electrons, which are accelerated in these regions
\cite{berezhkocasa}. Recently, the HEGRA experiments has detected an excess of
high-energy $\gamma$-rays from the supernova remnant Cas A, see \fref{casa}.
This is interpreted as evidence for hadron acceleration in the SNR.  The
hadrons interact with protons of the interstellar medium, producing $\pi^0$s,
which decay into high-energy photons, supposedly detected by the HEGRA
experiment \cite{berezhkocasa}. The flux is compatible with a model of electron
and hadron acceleration in shock fronts, see \fref{casa}.

Recently, an excess of charged particles from the direction of a SNR (Monogem
ring, $d\approx300$~pc) has been reported \cite{chilingarian}. However, such a
signal could not be confirmed by the KASCADE experiment \cite{kascadepoint}.
The latter has performed a detailed search for point sources, covering the
whole visible sky at energies $E_0>0.3$~PeV.  Special attention has been given
to the region of the galactic plane, as well as to the vicinity of known SNRs
and TeV-$\gamma$-ray sources.  No significant excess could be found. The
gyromagnetic radius\footnote{$r_G=p/(ZeB)$ for a particle with momentum $p$ and
charge $Ze$ in a magnetic field $B$.} of particles with an energy around 1~PeV
in the galactic magnetic field is in the order of 1~pc. Hence, it is not
expected to find any point sources at these energies.

\begin{figure}
 \includegraphics[width=0.5\columnwidth]{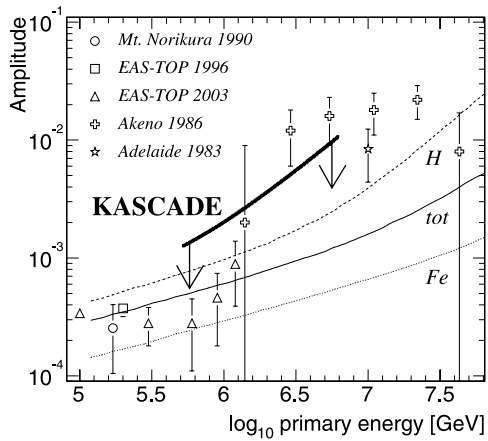} 
 \hspace*{\fill}
 \begin{minipage}[b]{0.45\textwidth}
 \caption{Rayleigh amplitudes as function of energy for various experiments,
   for references see \cite{kascadeaniso}. Additionally, model predictions for
   a diffusion model are shown. The lines indicate the expected anisotropy for
   primary protons (H) and iron nuclei (Fe) as well as for all particles (tot) 
   \cite{candiaaniso}.}
 \label{aniso}
 \end{minipage} \hh
\end{figure}
To characterize the large scale anisotropy of cosmic rays, the Rayleigh
amplitude$$A=\sqrt{\left(\frac{2}{n}\sum_{i=1}^n\sin\alpha_i\right)^2 
                 + \left(\frac{2}{n}\sum_{i=1}^n\cos\alpha_i\right)^2}$$
for $n$ events which are detected at a right ascension angle $\alpha_i$ is
introduced. A compilation of measured amplitudes as function of energy is
given in \fref{aniso} \cite{kascadeaniso}.  An increase of $A$ as function of
energy can be recognized. This trend is compatible with expectations taking
into account diffusive propagation of cosmic rays in the Galaxy
\cite{candiaaniso}, as indicated by the lines.

\section{Energy spectrum of cosmic rays}
\begin{figure}[b]
 \vspace*{-4mm}
 \includegraphics[width=0.59\columnwidth]{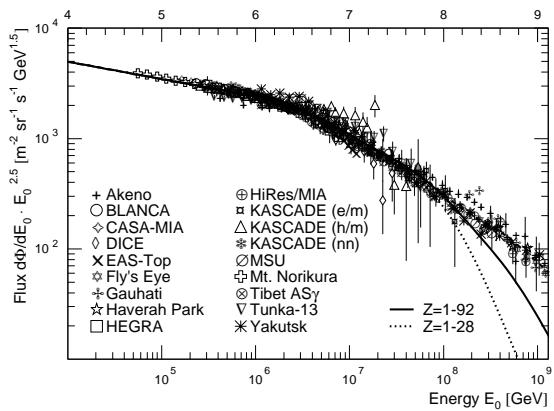}
 \begin{minipage}[b]{0.4\columnwidth}
   \caption{Normalized all-particle energy spectra for individual experiments,
   for details and references see \cite{knie}.}
   \label{knie1} 
 \end{minipage} \hh
\end{figure}
The all-particle energy spectrum has been derived by many experiments as shown
in \fref{knie1}. In this representation the flux of the air shower experiments
has been normalized to the extrapolated flux of direct measurements at 1~PeV by
introducing a slight adjustment ($\pm10\%$) of the energy scales of the
individual experiments \cite{knie}.  It is quite interesting to realize that
the absolute energy calibration of these various experiments, using different
observation techniques and models to describe the shower development in the
atmosphere agree within $\pm10\%$.  The normalized all-particle flux changes
smoothly without any prominent structures.  The solid and doted lines indicate
the total sum of galactic cosmic rays according to a parameterization of data
\cite{knie}.

\begin{figure} \centering
\includegraphics[width=0.49\columnwidth]{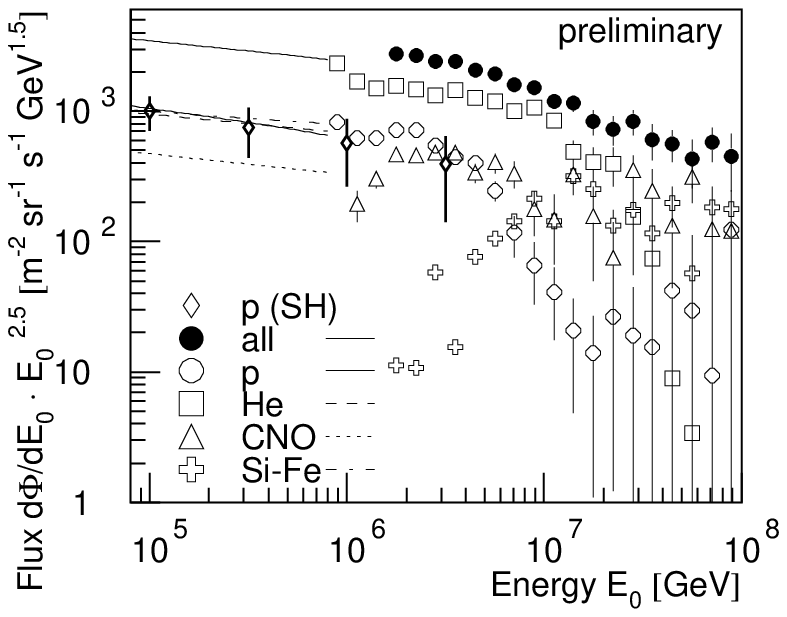}
\includegraphics[width=0.49\columnwidth]{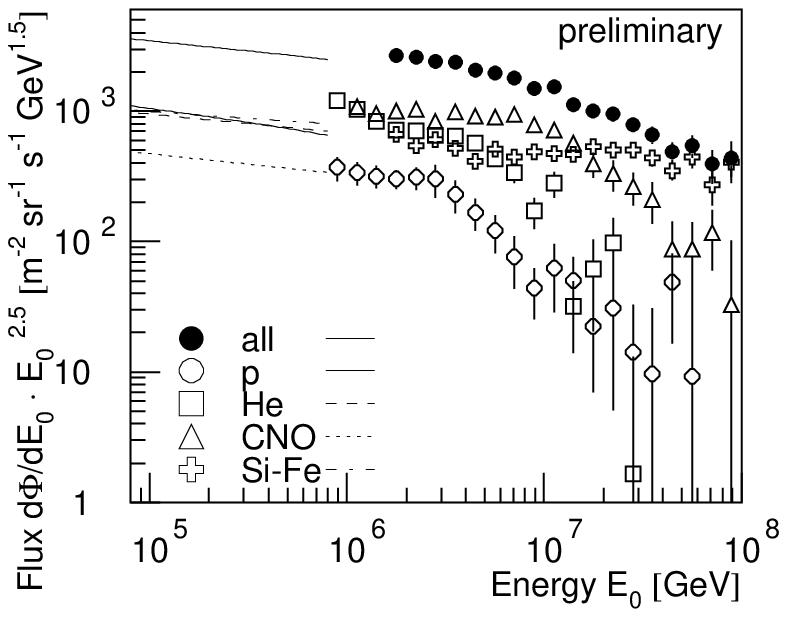}
\caption{Results from the KASCADE experiment: Energy spectra for groups of
  elements derived from the data using CORSIKA with the hadronic interaction
  models QGSJET (left) and SIBYLL (right) \cite{kascade}.  In addition, the
  proton spectrum as derived from the measurements of single hadrons is shown
  \cite{shspek}.  The lines indicate extrapolations of direct measurements
  \cite{knie}.}
\label{kascade} \hh
\end{figure}

\begin{figure}
 \includegraphics[width=0.5\columnwidth]{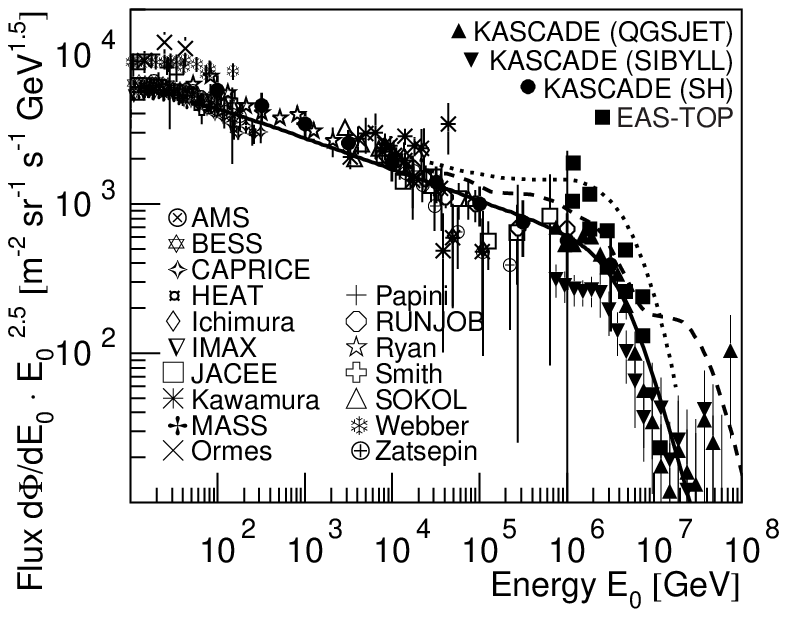}%
 \includegraphics[width=0.5\columnwidth]{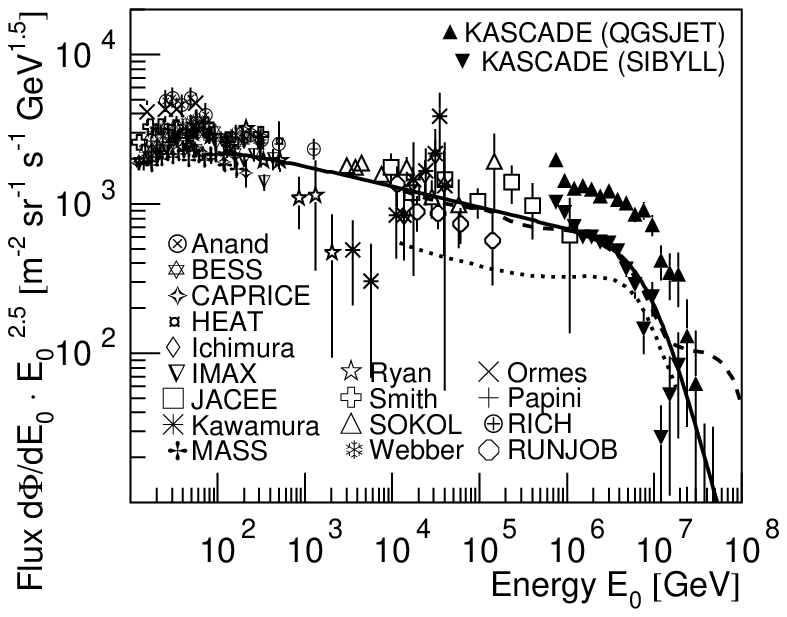}
 \includegraphics[width=0.5\columnwidth]{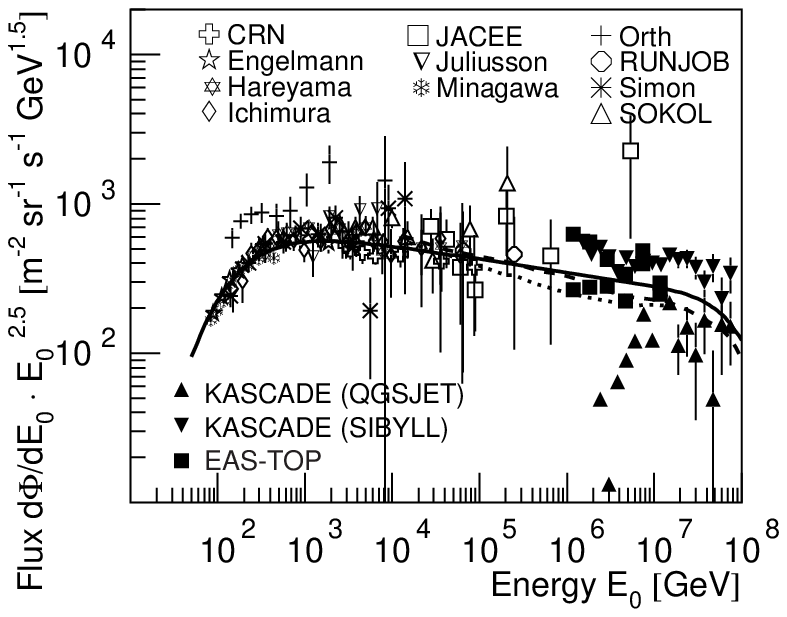}
 \begin{minipage}[b]{0.49\columnwidth}
 \caption{Energy spectra for elemental groups. Top left: protons,
  top right: helium, and bottom: iron nuclei.  Open symbols give results of
  direct measurements, for references see \cite{knie}. Filled symbols represent
  results from air shower measurements: KASCADE electrons/muons interpreted
  with two interaction models \cite{kascade} (preliminary), KASCADE single
  hadrons \cite{shspek}, and EAS-Top \cite{eastop}.   
  The data are compared to
  predictions of calculations by Kalmykov \etal\ \cite{kalmykov} (\dotted),
  Sveshnikova \etal\ \cite{sveshnikova} (\dashed), and the Poly-Gonato model
 \cite{knie} (\line).}
 \label{elementspek} 
 \end{minipage} \hh
\end{figure}

The KASCADE group performed systematic studies to evaluate the influence of
different hadronic interaction models used in the simulations to interpret the
data on the resulting spectra for elemental groups \cite{kascade}.  Two sets of
spectra, derived from the observation of the electromagnetic and muonic air
shower components, applying the Gold algorithm and using CORSIKA \cite{corsika}
with the hadronic interaction models QGSJET and SIBYLL are compiled in
Figure~\ref{kascade}.  

As can be seen in the figure, the flux for elemental groups depends on the
model used.  The KASCADE group emphasizes that at present, there are systematic
differences between measured and simulated observables, which result in the
ambiguities of the spectra for elemental groups. These conclusions apply in a
similar way also to other experiments.  A correct deconvolution of energy
spectra for elemental groups requires a precise knowledge of the hadronic
interactions in the atmosphere.  The interaction models presently used do not
describe the measurements with a sufficiently high precision required for this
task.  

The figure also shows the spectrum of primary protons, which has been derived
from the flux of unaccompanied hadrons \cite{shspek}.  The spectrum is
compatible with the proton flux as obtained from the unfolding procedure when
using the QGSJET model.  For comparison, also the results of direct
measurements at lower energies at the top of the atmosphere \cite{knie} are
presented in the figure.

In order to give an impression of the present status, the energy spectra for
three elemental groups (protons, helium, and iron nuclei) are compiled in
\fref{elementspek} over a wide energy range. Shown are results from direct
measurements above the atmosphere as well as the KASCADE results described
above.  The EAS-TOP experiment published two sets of spectra with different
assumptions about the contribution of protons and helium nuclei, for details
see \cite{eastop}. The resulting fluxes are indicated by two squares per
primary energy.  To guide the eye, the solid lines indicate power law spectra
with a cut-off at $Z\cdot4.5$~PeV. For the iron spectrum at low energies the
influence of modulation due to the magnetic fields of the heliosphere can
be recognized.

The dashed lines represent calculations of energy spectra for nuclei
accelerated in supernova remnants by Sveshnikova \etal\ \cite{sveshnikova}.  It
is assumed that the particles are accelerated in a variety of supernovae
populations, each having an individual maximum energy to be attained during
acceleration, which results in the bumpy structure of the obtained spectra.
The dotted lines reflect calculations of the diffusive propagation of particles
through the Galaxy by Kalmykov \etal\ \cite{kalmykov}. The leakage of particles
yields a rigidity dependent cut-off.
Comparison with the data may suggest a {\sl qualitative} understanding of the
energy spectra. However, for a precise {\sl quantitative} understanding,
detailed investigations of the systematic errors of the measurements are
necessary and the description of the interaction processes in the atmosphere
needs to be improved.

\section{Mass composition of cosmic rays}
The position of the maximum of an electromagnetic or nuclear cascade in matter
depends on the incident particles energy as $X_{max}\propto\ln E_n$, where
$E_n$ is the energy per nucleon $E_n=E_0/A$. Hence, $X_{max}$ depends on $\ln
A$. Also, many other observables in air showers like the number of electrons,
muons, or hadrons observed at ground level depend roughly on $\ln A$.  To
characterize the cosmic-ray mass composition one uses commonly the mean
logarithmic mass $\lnA$, defined as $\lnA=\sum r_i\ln A_i$, where $r_i$ is the
relative fraction of nuclei with atomic mass number $A_i$.

The mean logarithmic mass from experiments measuring electrons, muons, and
hadrons at ground level using mostly CORSIKA/QGSJET to interpret the data are
compiled in \fref{masse}c. One recognizes an increase of $\lnA$ as function
of energy, the line indicates an increase as expected for a rigidity dependent
cut-off for individual elements \cite{knie}.

\begin{figure} 
 \ifnum \flag = 1 
 \begin{minipage}[b]{0.49\columnwidth}
 \includegraphics[width=\columnwidth]{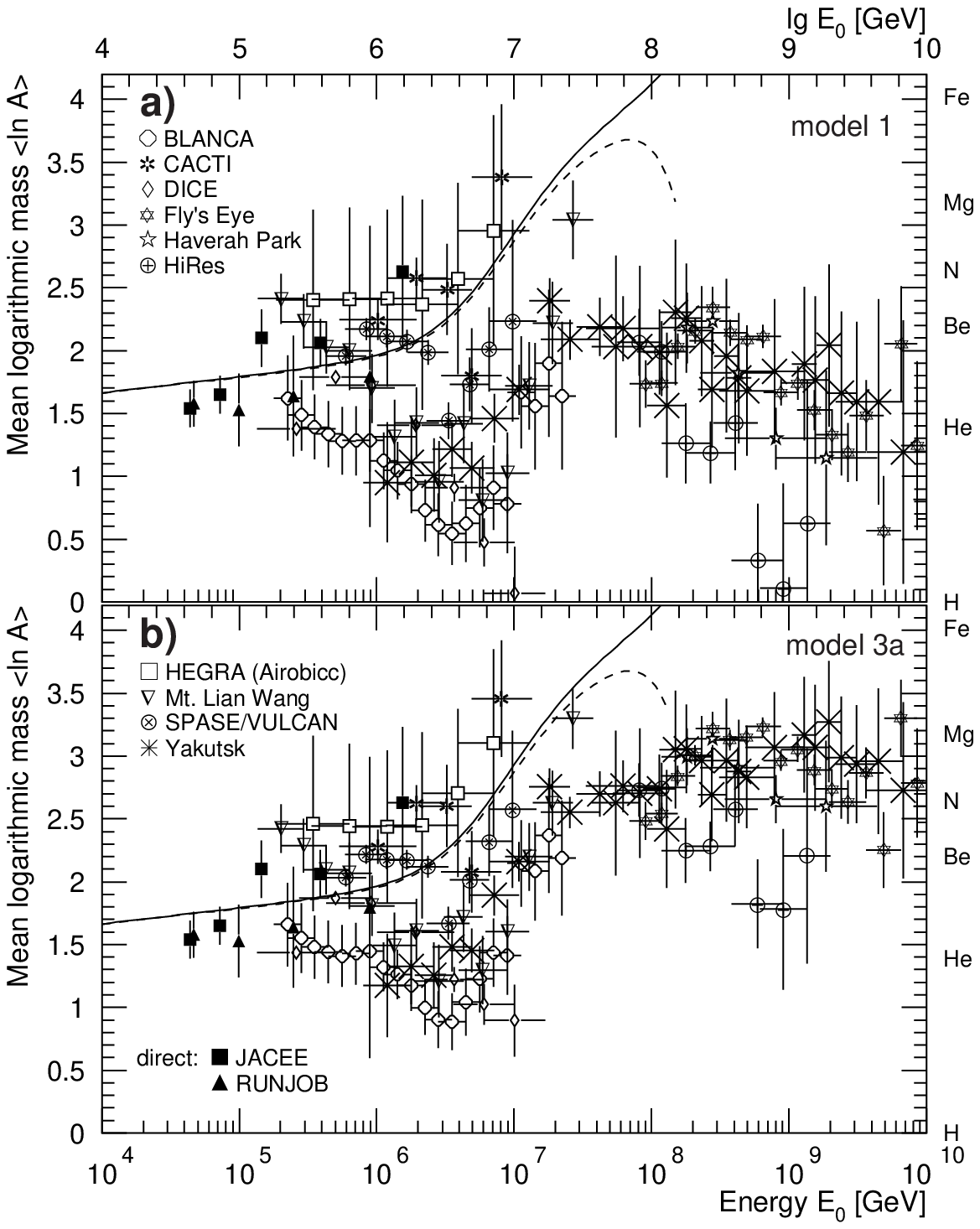} 
 \vspace*{0.1mm}
 \end{minipage}\hspace*{\fill}
 \begin{minipage}[b]{0.49\columnwidth}
 \includegraphics[width=\columnwidth]{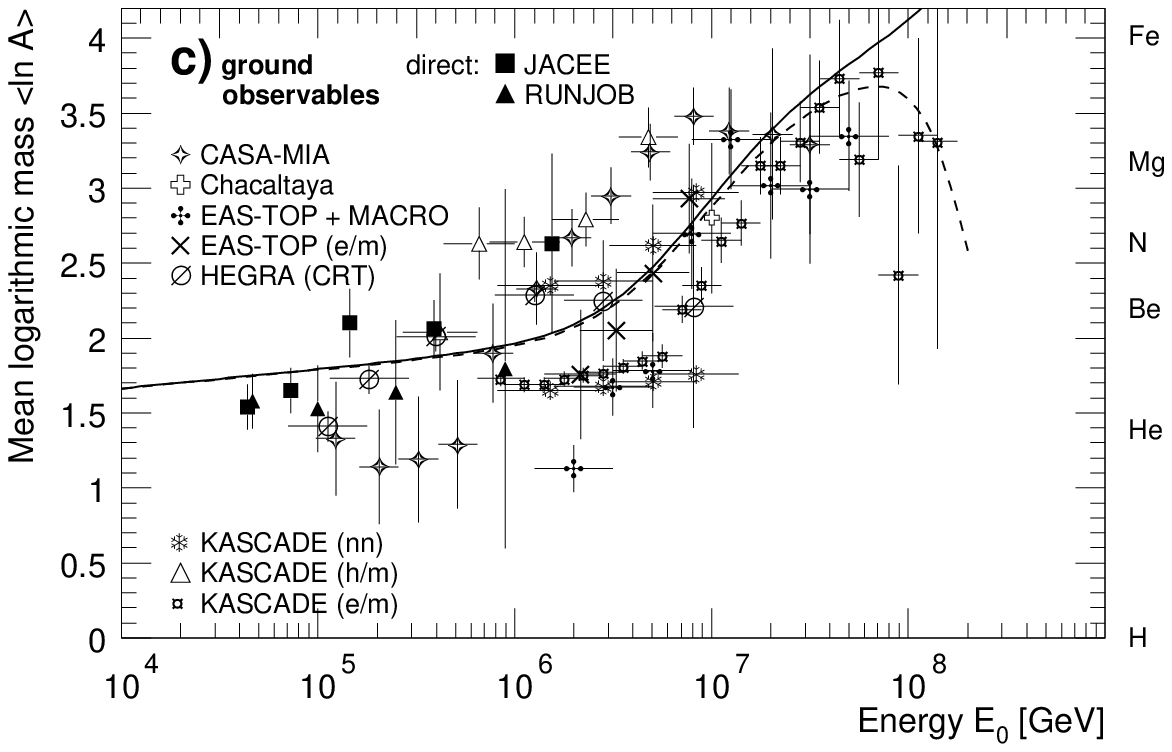}\\[-7mm]
 \caption{Mean logarithmic mass as function of energy as obtained by various
  experiments. 
  Left: $\lnA$ derived from observations of the average depth of the shower
  maximum $\Xmax$ interpreted with CORSIKA/QGSJET01 (top) and with a modified
  version (bottom) \cite{wq}.
  Right: Experiments measuring electrons, muons, and hadrons at ground level.
  For details and references see \cite{knie,wq}.}
  \label{masse} 
 \end{minipage} \hh
 \else 
 \centering
 \includegraphics[width=0.69\columnwidth]{hoerandel_2004_01_fig07a.eps} 
 \includegraphics[width=0.69\columnwidth]{hoerandel_2004_01_fig07b.eps}\\[-3mm]
 \caption{Mean logarithmic mass as function of energy as obtained by various
  experiments. 
  Top: $\lnA$ derived from observations of the average depth of the shower
  maximum $\Xmax$ interpreted with CORSIKA/QGSJET01 (a) and with a modified
  version (b) \cite{wq}.
  Bottom: Experiments measuring electrons, muons, and hadrons at ground level.
  For details and references see \cite{knie,wq}.}
  \label{masse} 
  \hh
 \fi 
\end{figure}

A second group of experiments measures the average depth of the shower maximum
$\Xmax$ in the atmosphere by registering \v{C}erenkov photons or fluorescence
light. Interpreting this data with CORSIKA/QGSJET~01 leads to the results shown
in \fref{masse}a. An increase of $\lnA$ as described before is not obtained.

The description of hadronic interactions within the models is based on
accelerator measurements. However, extrapolating the experimental uncertainties
to higher energies may cause significant differences in the interpretation of
air shower data. For example, the cross sections for proton - anti proton
collisions at the Tevatron $(\sqrt{s}=1.8$~TeV$)$ exhibit an error of about
10\%. To investigate the influence of such uncertainties on the development of
air showers, in the model QGSJET 01 the cross section for $p$-$\bar{p}$
collisions has been reduced to the lower error boundary of the Tevatron data.
This results in a deeper penetration of showers into the atmosphere.
Accordingly, $\lnA$ derived with such an interaction model, using the same
experimental $\Xmax$-values as before, yields larger values at high energies,
as shown in \fref{masse}b \cite{wq}. These values follow better the trend of
the observations of particles at the ground, shown in \fref{masse}c.

\begin{figure} \centering
 \includegraphics[width=0.46\columnwidth]{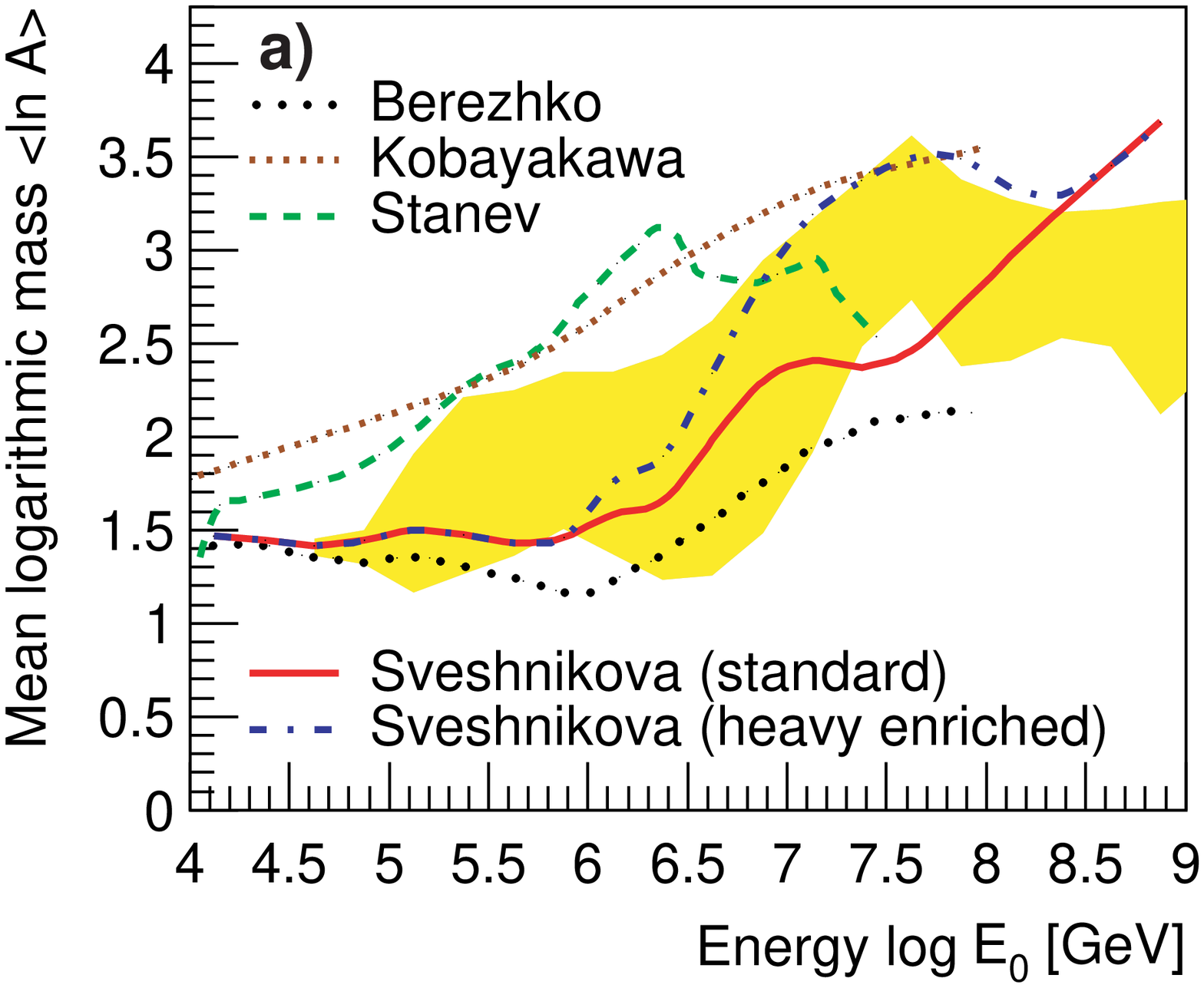}
 \includegraphics[width=0.46\columnwidth]{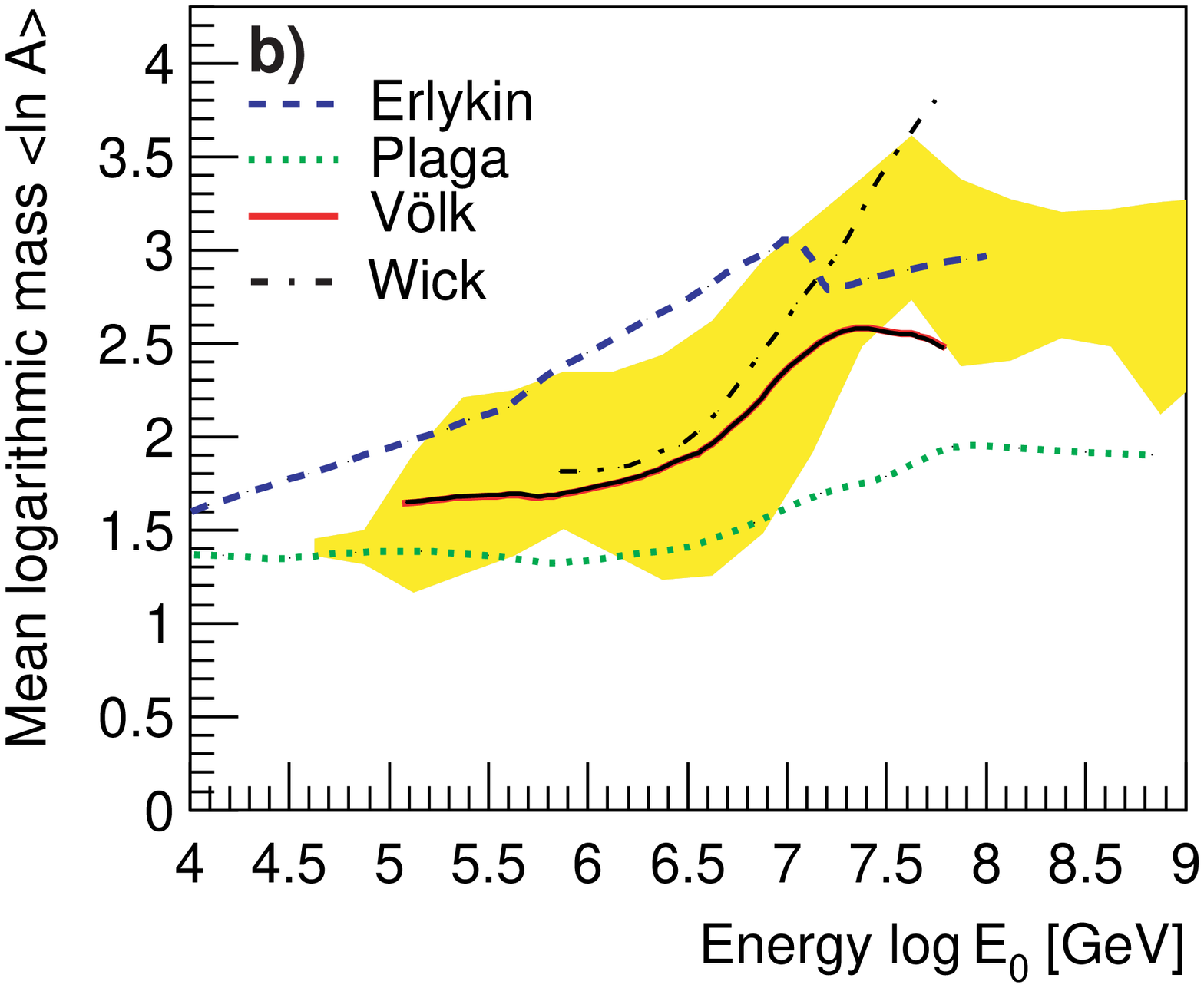}
 \includegraphics[width=0.46\columnwidth]{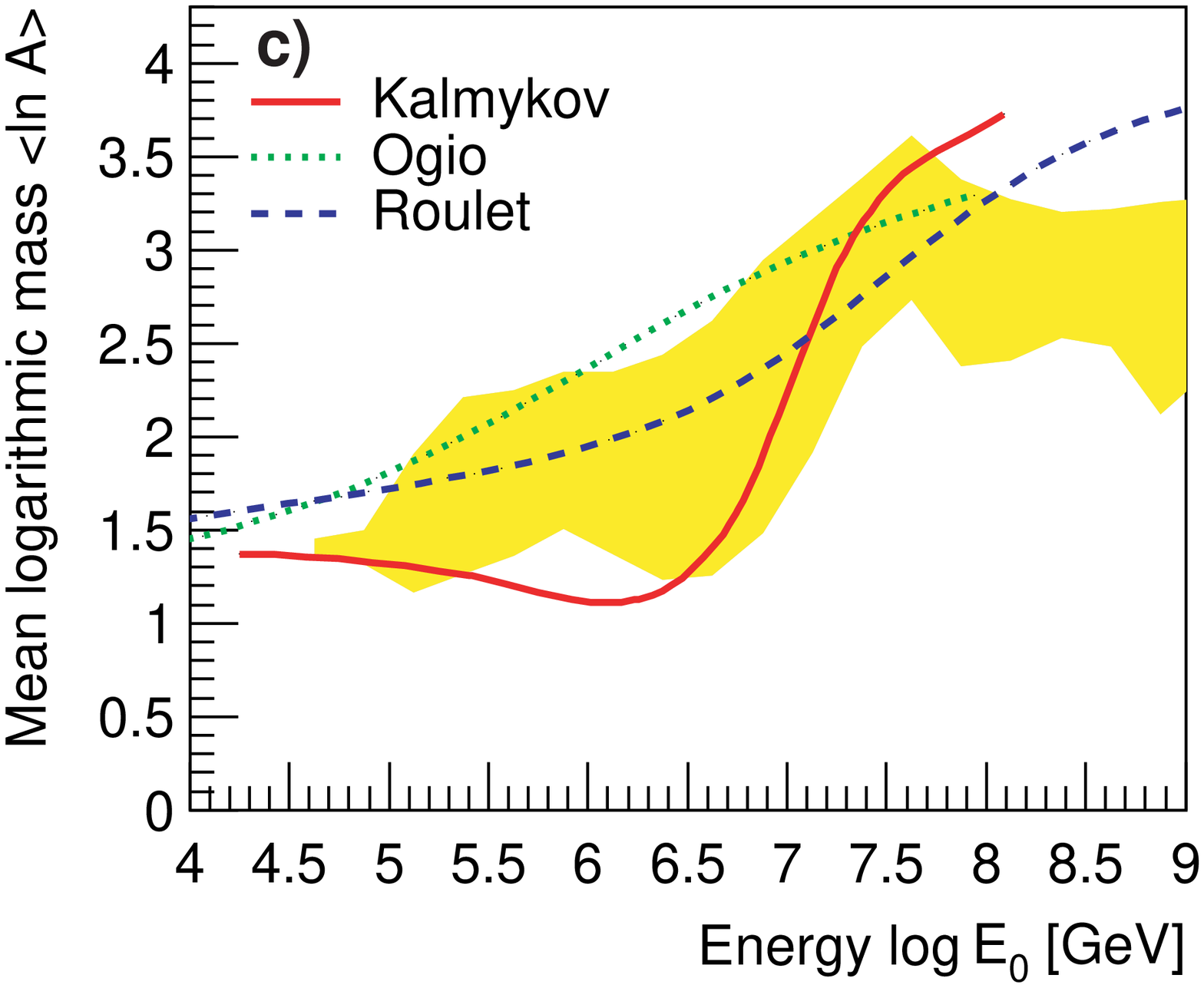}
 \includegraphics[width=0.46\columnwidth]{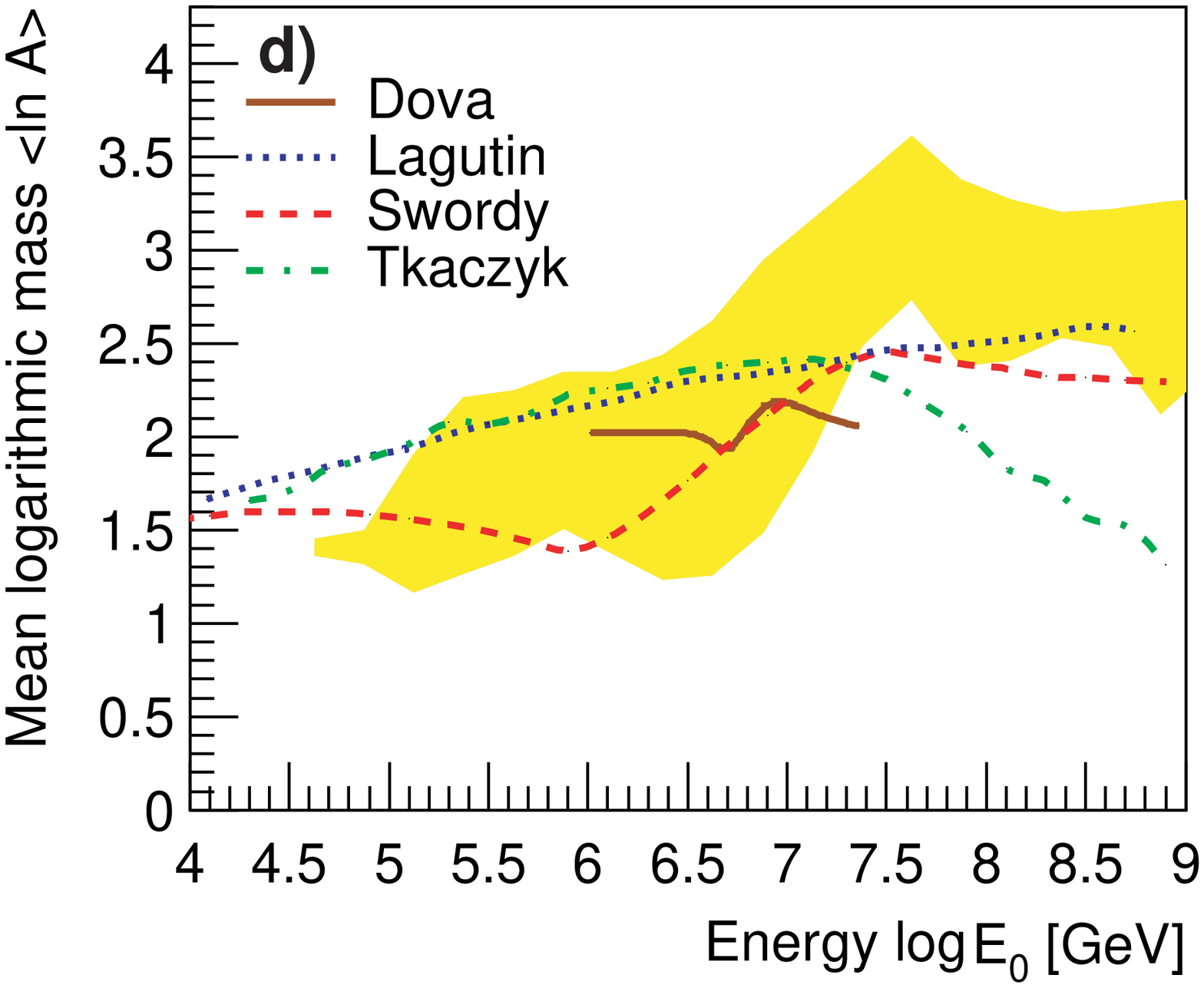}
 \caption{Mean logarithmic mass as function of energy as obtained by various
  experiments (shaded area) compared with different models (lines).
  a) Acceleration in SNR (Berezhko \etal\ \cite{berezhko}, Stanev \etal\
     \cite{stanev}, Kobayakawa \etal\ \cite{kobayakawa}, Sveshnikova \etal\
     \cite{sveshnikova});
  b) Single source model (Erlykin \& Wolfendale \cite{wolfendale}),
    acceleration in $\gamma$-ray bursts (Plaga \cite{plaga}, Wick \etal\
    \cite{wick}), reacceleration in the galactic wind (V\"olk \etal\
    \cite{voelk});
  c) Diffusion in Galaxy (Kalmykov \etal\ \cite{kalmykov}, Ogio \etal\
     \cite{ogio}, Roulet \etal\ \cite{roulet});
  d) Propagation in the Galaxy (Lagutin \etal\ \cite{lagutin}, 
     Swordy \cite{swordy}) and interaction with background photons (Tkaczyk
     \cite{tkaczyk}) and neutrinos (Dova \etal\ \cite{dova}).
  For details see \cite{origin}.}
 \label{lna} \hh
\end{figure}

Combining the experimental values from \fref{masse}b and c, the result is
displayed as grey band in \fref{lna}. The experimental data are going to be
compared to various models, which are discussed in the literature as possible
origin of the knee \cite{origin}.

Several approaches to model the shock acceleration in SNRs (Berezhko \etal\
\cite{berezhko}, Kobayakawa \etal\ \cite{kobayakawa}, Stanev \etal\
\cite{stanev}, Sveshnikova \etal\ \cite{sveshnikova}) are summarized in
\fref{lna}a. The models differ in assumptions of properties of the SNRs like
magnetic field strength, available energy etc.  This yields differences in
$\lnA$, as can be inferred from the figure.

Predictions of the single source model (Erlykin \& Wolfendale
\cite{wolfendale}), reacceleration in the galactic wind (V\"olk \etal\
\cite{voelk}) and acceleration in $\gamma$-ray bursts (Plaga \cite{plaga}, Wick
\etal\ \cite{wick}) are shown in \fref{lna}b. The latter differ in their
interpretation of the origin for the knee, Plaga attributes it to the maximum
energy reached during the acceleration process, while Wick \etal\ propose
leakage from the Galaxy as the cause.

The propagation of cosmic rays as described in diffusion models (Kalmykov
\etal\ \cite{kalmykov}, Ogio \etal\ \cite{ogio}, Roulet \etal\ \cite{roulet})
yields $\lnA$-values presented in \fref{lna}c. They are all based on the
approach by Ptuskin \etal\ \cite{ptuskin}, but take into account different
assumptions on details of the propagation process, like the structure of
galactic magnetic fields etc.

The last panel (\fref{lna}d) summarizes predictions of models taking into
account cosmic-ray propagation in the Galaxy (Lagutin \cite{lagutin}, Swordy
\cite{swordy}) as well as interactions with background photons (Tkaczyk
\cite{tkaczyk}) and neutrinos (Dova \etal\ \cite{dova}).

As can be inferred from \fref{lna}, the situation is not yet conclusive, but a
trend towards a standard picture can be recognized.  Some of the proposed
explanations can already be excluded.  Interactions with background particles
in the Galaxy would produce a big amount of secondary protons, which results in
a light mass composition at high energies, not confirmed by the experiments.
Furthermore, a massive neutrino, proposed in \cite{dova} is excluded by
measurements of the WMAP and 2dFGRS experiments \cite{hannestad}.  The approach
described in \cite{plaga} does not describe the trend of the data.  The data are
consistent with acceleration in SNRs (\fref{lna}a) and diffusive propagation
(\fref{lna}c). Also reacceleration in the galactic wind (V\"olk \etal) and
acceleration in $\gamma$-ray bursts in combination with diffusive propagation
(Wick \etal) follow the trend indicated by the data.

\section{Conclusions}
Comparing the present results to the status before the start of the KASCADE
experiment (about one decade ago), our knowledge about high-energy cosmic rays
has significantly improved. The experiment has shown that the knee is caused by
the subsequent cut-offs of individual elements, starting with  protons and
helium nuclei and that the mean logarithmic mass increases as function of
energy.

Summarizing the large number of experimental observations, there are
indications for a standard picture of galactic cosmic rays. At least a large
fraction of them seems to be accelerated in supernova remnants up to energies
of $Z\cdot(0.1-5)$~PeV. Higher energies may be reached by reacceleration in the
galactic wind or by acceleration in additional sources, such as $\gamma$-ray
bursts. The particles propagate in a diffusive process through the Galaxy.
With rising energy the pathlength decreases and particles escape easier from
the Galaxy. This brings about the knee in the energy spectrum.

When the understanding of the hadronic interactions in the atmosphere improves,
the measurements can be interpreted with higher reliability. This will put more
restrictions on the models to describe acceleration and propagation of cosmic
rays.

\section*{Acknowledgments}
The author acknowledges valuable scientific discussions with his colleagues
from the KASCADE-Grande and TRACER experiments. He thanks the organizers of the
Vulcano workshop for their invitation, the great hospitality, and the
interesting scientific program.

%

{\footnotesize
\begin{thebibliography}{99}
\itemsep 0.0mm
\bibitem{kulikov} G.V.~Kulikov \etal, J. Exp. Theor. Phys. 35 (1958) 635
\bibitem{knie} J.R. H\"orandel, Astropart. Phys. 19 (2003) 193
\bibitem{atic} T.G.~Guzik \etal, Adv. Space Res., in press 2004
\bibitem{tracer} F.~Gahbauer \etal, ApJ 607 (2004) 333
\bibitem{cerntrd} S.P. Wakely \etal, Nucl. Instr. and Meth., in press (2004)
\bibitem{hadron} J.R.~H\"orandel, Nucl. Phys. B (Proc. Suppl.) 122 (2003) 455
\bibitem{baade} W. Baade and F. Zwicky, Phys. Rev. 46 (1934) 76
\bibitem{fermi} E. Fermi, Phys. Rev. 75 (1949) 1169 
\bibitem{blanford} R.D. Blanford and J.P. Ostriker, ApJ 221 (1978) L29
\bibitem{berezhko} E.G. Berezhko and L.T. Ksenofontov, JETP 89, 3 (1999) 391
\bibitem{stanev} T. Stanev et al., Astron. \& Astroph. 274 (1993) 902
\bibitem{kobayakawa} K. Kobayakawa et al., Phys. Rev. D 66 (2002) 083004
                     and preprint astro-ph/0008209
\bibitem{sveshnikova} L.G. Sveshnikova, Astron. \& Astroph., 409 (2003) 799
\bibitem{wolfendale} A.D. Erlykin and A.W. Wolfendale, 
                  J. Phys. G: Nucl. Part. Phys. 27 (2001) 1005
\bibitem{plaga} R. Plaga, New Astronomy 7 (2002) 317
\bibitem{wick} S.D. Wick \etal, Astropart. Phys. 21 (2004) 125
\bibitem{ptuskin} V.S. Ptuskin et al., Astron. \& Astroph. 268 (1993) 726.
\bibitem{ogio}    S. Ogio and F. Kakimoto, Proc. 28th Int. Cosmic Ray Conf.,
                  Tsukuba 1 (2003) 315
\bibitem{roulet}  R. Roulet, preprint astro-ph/0310367
\bibitem{swordy} S.P. Swordy, Proc. 24th Int. Cosmic Ray Conf., 
                Rome 2 (1995) 697
\bibitem{lagutin} A.A. Lagutin et al., Nucl. Phys. B (Proc. Suppl.) 97 (2001) 
                  267
\bibitem{voelk}  H.J. V\"olk and V.N. Zirakashvili, 
                 Proc. 28th Int. Cosmic Ray Conf., Tsukuba 4 (2003) 2031
\bibitem{tkaczyk} S. Karakula and W. Tkaczyk, Astropart. Phys. 1 (1993) 229
\bibitem{dova} M.T. Dova et al., preprint astro-ph/0112191
\bibitem{candia} J. Candia et al., Astropart. Phys. 17 (2002) 23
\bibitem{kazanas} D. Kazanas and A. Nicolaidis, preprint astro-ph/0103147
\bibitem{origin} J.R. H\"orandel, Astropart. Phys. 21 (2004) 241
\bibitem{hegra} F.~Aharonian \etal, Astron. \& Astroph. 370 (2001) 112
\bibitem{berezhkocasa} E.G.~Berezhko \etal, Astron. \& Astroph. 400 (2003) 971
\bibitem{chilingarian} A. Chilingarian \etal, ApJ 597 (2003) L129
\bibitem{kascadepoint} Kascade Coll.: T. Antoni \etal, ApJ 608 (2004) 865
\bibitem{kascadeaniso} Kascade Coll.: T. Antoni \etal, ApJ 604 (2004) 687
\bibitem{candiaaniso} J. Candia \etal, J. Cosmol. Astropart. Phys., 5 (2003) 3
\bibitem{kascade} KASCADE Coll.:
 H. Ulrich \etal, European Physical Journal C (2004) DOI:
10.1140/epjcd/s2004-03-1632-2; K.-H. Kampert \etal, Acta Physica Polonica B 35
(2004) 1799; J.R. H\"orandel \etal, preprint aspro-ph/0311478.
\bibitem{shspek} KASCADE Coll.: T.~Antoni \etal, ApJ, in press (2004);
M. M\"uller \etal, Proc. 28th Int. Cosmic Ray Conf., Tsukuba 1 (2003) 101
\bibitem{eastop}  G. Navarra et al., Proc. 28th Int. Cosmic Ray Conf.,
                  Tsukuba 1 (2003) 147;
                  S. Valchierotti et al., Proc. 28th Int. Cosmic Ray Conf.,
                  Tsukuba 1 (2003) 151
\bibitem{corsika} D. Heck et al., Report FZKA 6019, 
         Forschungszentrum Karlsruhe 1998;\\ 
         and http://www-ik.fzk.de/$\sim$heck/corsika.
\bibitem{kalmykov} N.N. Kalmykov and A.I. Pavlov, Proc. 26th Int. Cosmic
                  Ray Conf., Salt Lake City 4 (1999) 263
\bibitem{wq} J.R. H\"orandel, J. Phys. G: Nucl. Part. Phys 29 (2003) 2439
\bibitem{hannestad} S. Hannestad, preprint astro-ph/0303076
\end{thebibliography}
}
\end{document}